\documentclass[a4paper,11pt]{article}
\usepackage{jheppub} 
\usepackage{lineno}
\usepackage{enumitem}
\usepackage{cleveref}
\usepackage{bm}
\usepackage[percent]{overpic}
\usepackage{latexsym}
\usepackage{dcolumn}
\usepackage{amsmath,amsfonts,amssymb}
\usepackage{graphicx,epsfig}
\usepackage{psfrag}
\usepackage{amsmath,amssymb}
\usepackage{amsthm}
\usepackage{color}
\usepackage{wasysym,breqn}
\usepackage{natbib}
\usepackage{float}
\usepackage{moresize}
\usepackage{enumerate}
\usepackage[hang,flushmargin]{footmisc}
\usepackage[titletoc,toc]{appendix}
\usepackage{lipsum}
\usepackage{subcaption}
\usepackage{epstopdf} 
\usepackage{bookmark}
\usepackage{rotating}
\usepackage{multirow}
\usepackage{multicol}
\usepackage{mathtools}
\usepackage[font=small,labelfont=bf,
   justification=justified,
   format=plain]{caption}
\usepackage{tikz}
\usepackage{stackengine}
\usepackage{scalerel}
\usepackage{tabularray}
\usepackage{xtab,afterpage}
\usepackage{siunitx}

\title{Power-Law Bounces in $f(R)$ Gravity: Analysis of the Ekpyrosis and Accelerating Regimes}
\author[a]{Saurya Das,}
\author[b, d, e]{Peter Dunsby,}
\author[b, c]{S. Shajidul Haque}
\author[b]{and Seturumane Tema}

\affiliation[a]{Department of Physics and Astronomy \& Quantum Alberta, University of Lethbridge, 4401 University Drive, Lethbridge AB T1K 3M4, Canada}
\affiliation[b]{Department of Mathematics and Applied Mathematics, University of Cape Town,\\ Cape Town-7700, South Africa}
\affiliation[c]{National Institute for Theoretical and Computational Sciences (NITheCS),\\ Private Bag X1, Matieland,
South Africa}
\affiliation[d]{South African Astronomical Observatory, Observatory 7925, Cape Town, South Africa}
\affiliation[e]{Centre for Space Research, North-West University, Potchefstroom 2520, South Africa}
\emailAdd{saurya.das@uleth.ca}
\emailAdd{peter.dunsby@uct.ac.za}
\emailAdd{shajid.haque@uct.ac.za}
\emailAdd{tmxset001@myuct.ac.za}

\abstract{We investigate the dynamics of the Friedmann-Lemaître-Robertson-Walker spacetime within the framework of $f(R)$ gravity using a compact, model-independent dynamical systems approach. By assuming a power-law scale factor, 
we explore ekpyrotic 
and accelerating 
solutions to address the big bang singularity. Our analysis demonstrates that a cosmological bounce, characterized by a transition from contraction to expansion, possibly avoids the singularity without directly using the Raychaudhuri equation, unlike previous approaches using specific \( f(R) \simeq R^n \) forms. We identify a key fixed point in the phase space corresponding to the bounce, supported by perturbation analysis and qualitative description of trajectories in the phase space. The results suggest that \( f(R) \) gravity provides a robust framework for non-singular cosmologies.
\\
}

\begin{document}

\maketitle

\section{Introduction}\label{Introduction}
The standard $\Lambda$CDM model of cosmology has achieved remarkable success in explaining a wide range of observations, including the cosmic microwave background anisotropies, large-scale structure, and type Ia supernovae \cite{hawking2023large,peebles2020principles,ostriker1995cosmic}. However, it predicts a big bang singularity approximately 13.79 billion years ago, where the universe's scale factor vanishes, while its temperature, and density diverge, and time-like geodesics terminate \cite{senovilla1998singularity, hawking1970singularities,penrose1965gravitational,senovilla2022critical}. Resolving this singularity is a critical challenge in cosmology, as it questions the physical validity of initial conditions in the early universe.
Modified gravity theories, such as $f(R)$ gravity, offer a promising avenue to resolve singularities by altering the gravitational dynamics \cite{chakraborty2024model, das2024cosmological,arora2022can}. In \( f(R) \) gravity, the Einstein-Hilbert action is generalized by replacing the Ricci scalar \( R \) with an arbitrary function \( f(R) \), allowing for non-standard cosmological evolutions, including bounces \cite{starobinsky2007disappearing,libanov2016generalized,bamba2014bounce}. A cosmological bounce is a non-singular transition from a contracting phase (\(\dot{a} < 0\)) to an expanding phase (\(\dot{a} > 0\)), maintaining a finite scale factor \cite{buchbinder2007new}. Such scenarios are prevalent in ekpyrotic and cyclic cosmologies, where a slow contraction (ekpyrosis) precedes the bounce, potentially driven by scalar fields or modified gravity effects \cite{arora2022can,lehners2018small}. \\

Previous studies have addressed the big bang singularity in \(f(R) \simeq R^n \) gravity by demonstrating that a violation of the convergence condition in the Raychaudhuri equation can lead to an evasion of classical singularity theorems \cite{hawking1970singularities,penrose2019singularity,hawking1966occurrence,ellis2003closed}. More recently, a different approach has been proposed based on a model-independent dynamical system framework in \( f(R) \) gravity \cite{chakraborty2024model, chakraborty2021model,carloni2009some}. This method employs compact phase space variables constructed from cosmographic parameters, enabling a general analysis without specifying the functional form of \( f(R) \), making it ideal for studying various solutions of the scale factor to investigate cosmological bounces that avoid the initial singularity.\\ 

In this work, we use this compact, model-independent dynamical systems approach to explore cosmological bounces in $f(R)$ gravity, focusing on power-law solutions. By analyzing the accelerating and ekpyrotic regimes, we identify a key fixed point in the phase space corresponding to a non-singular bounce, where the universe transitions from contraction to expansion with a finite minimum scale factor. Our perturbation analysis and phase space trajectories confirm the stability and dynamics of this bounce, demonstrating that $f(R)$ gravity provides a robust framework for tackling the big bang singularity without using specific forms of $f(R)$.
These results generalize previous findings and suggest potential extensions to other modified gravity theories, such as Brans-Dicke theory, offering new insights into non-singular cosmologies.\\ 

The rest of the paper is organized as follows. 
Section \ref{sec:power_law} introduces the power-law solution and cosmographic parameters. Section \ref{sec:dynamical_system} presents the compact dynamical system formulation, fixed points, and their cosmological implications. Section \ref{sec:bounce} analyzes the bounce at a key fixed point, supported by perturbation and trajectory analysis. Section \ref{sec:conclusion} summarizes the findings and discusses future directions.

\section{Cosmography and Power-law Solutions}\label{sec:power_law}
In this paper, we utilize a compact dynamical systems formulation of $f(R)$ gravity, based on cosmography, which involves expanding the scale factor as a Taylor series in cosmic time \cite{chakraborty2024model,capozziello2008cosmography,dunsby2016theory}. This expansion yields various parameters, including the Hubble parameter, deceleration, jerk, snap, lerk, and higher-order terms. The number of cosmographic parameters required depends on the specific problem under investigation. Typically, cosmography is applied to late-time cosmology, with general models considering an expansion up to the fifth order, involving five parameters \cite{dunsby2016theory}. However, this approach is not inherently limited to late-time scenarios.\\

Here we focus on early-time cosmology, exploring power-law solutions of the form 
\begin{equation}\label{laqq}
a(t) \propto t^\beta\;,
\end{equation}
where $\beta$ is a constant. Specifically, we investigate ekpyrotic ($\beta < \frac{1}{3}$) and accelerating ($\beta > 1$) regimes to identify cosmological bounce scenarios that potentially avoid the initial singularity. These power-law solutions and their implications for non-singular cosmologies have been explored in various modified gravity frameworks, including gauged supergravity \cite{Blaback:2013sda}, as well as in studies addressing consistent embeddings of cosmic bubbles \cite{haque2017consistent} and extensions of the Raychaudhuri equation in modified gravity \cite{burger2018towards}.\\

It was shown in \cite{chakraborty2024model} that a model-independent compact dynamical systems formulation of $f(R)$ gravity requires only three parameters $H$, $q$, and $j$ corresponding to the first, second, and third orders of the expansion, respectively. Below, we provide an overview of power-law solutions, followed by the definitions of the relevant cosmographic parameters.\\

The continuity equation in general relativity relates the energy density $\rho$ to the scale factor via $\rho \propto a^{-3(w+1)}$, where $w$ is the equation of state parameter, which takes the values $w = 0$ for dust matter and $w = \frac{1}{3}$ for radiation. In this framework, the universe evolves through a radiation-dominated epoch followed by a matter-dominated epoch. During these epochs, the power $\beta$ takes the values $\beta = \frac{1}{2}$ and $\beta = \frac{2}{3}$, respectively, with the Hubble parameter $H = \frac{\dot{a}}{a}$ being positive, indicating an expanding universe. As $t \to 0^+$, the scale factor in (\ref{laqq}) approaches zero, causing the densities of radiation and matter to diverge. To eliminate these infinities, we assume a finite scale factor when $t \to 0$. This scenario implies the existence of a $t \to 0^-$ epoch, known as ekpyrosis \cite{buchbinder2007new, steinhardt2002cosmic, brandenberger2020reheating}, where the scale factor follows (\ref{laqq}) in the negative time direction ($t < 0$) with $\beta < \frac{1}{3}$. During this ekpyrotic phase, the Hubble parameter $H$ is negative, indicating a slow contraction to a minimum finite scale factor compared to the contraction obtained by reversing time in radiation or matter-dominated epochs. This contraction, as already mentioned in section \ref{Introduction}, is typically driven by matter in the form of a scalar field or a stiff fluid \cite{ijjas2018bouncing, chavanis2015cosmology}. 
\\

Next, we define the deceleration ($q$) and jerk ($j$) parameters in terms of the second and third derivatives of (\ref{laqq}):
\begin{flalign}
\label{acv}
\begin{aligned}
q &= -1 - \frac{\dot{H}}{H^2} = -1 + \frac{1}{\beta}, \\
j &= -2 - 3q + \frac{\ddot{H}}{H^3} = -2 - 3q + \frac{2}{\beta^2}.
\end{aligned}
\end{flalign}
The deceleration parameter $q$ quantifies the rate of change of the universe's expansion. During the radiation and matter-dominated epochs, $q > 0$ indicates deceleration due to the presence of matter slowing the expansion. For $\beta > 1$ in (\ref{laqq}), the deceleration parameter is negative ($q < 0$), corresponding to an accelerating universe. Such power-law solutions with $\beta > 1$ are termed accelerating solutions, typically driven by matter with an equation of state parameter $w = -1$. 
The parameters $q$ and $j$ are crucial in formulating a compact, model-independent dynamical systems approach for $f(R)$ gravity \cite{chakraborty2024model, capozziello2008cosmography, macdevette2024model, chakraborty2022note}, which is introduced in the next section.


\section{Model Independent Compact Dynamical Formulation of $f(R)$ Gravity} \label{sec:dynamical_system}
To investigate non-singular cosmological bounces in $f(R)$ gravity, we draw on a model-independent dynamical systems framework developed in \cite{chakraborty2024model,arora2022can,chakraborty2021model}. This approach employs compact phase space variables to capture transitions from contraction to expansion at $H = \frac{\dot{a}}{a} = 0$ in FLRW spacetimes. Autonomous differential equations, independent of specific $f(R)$ forms, identify fixed points corresponding to bounces, with stability assessed via perturbation analysis. 
This framework, which can potentially be extended to other theories such as Brans–Dicke theory, guides our analysis of ekpyrotic and accelerating regimes without relying solely on the repulsive terms induced in the Raychaudhuri equation and their implications for the congruence of geodesics, as was done previously
%
for $f(R) \simeq R^n$ \citep{das2024cosmological}. In $f(R)$ gravity, the action is 
\begin{equation}\label{ens}
S = \frac{1}{2\kappa}\int d^4 x \sqrt{-g}\,f(R) + S_m,
\end{equation}
where $f(R)$ is an arbitrary function of the Ricci scalar, $\kappa=8\pi G$ and $S_m$ represents the matter action. This action generalizes the Einstein-Hilbert action by incorporating an arbitrary function $f(R)$. The field equations obtained by   varying (\ref{ens}) with respect to the metric tensor are 
\begin{equation} \label{gen_fieldeq}
F(R) R_{\mu\nu} - \frac{1}{2} f(R) g_{\mu\nu} - \nabla_\mu \nabla_\nu F(R) + g_{\mu\nu} \Box F(R) = \kappa T_{\mu\nu}\;.
\end{equation}
These fourth-order field equations govern the dynamics of $f(R)$ gravity. When one imposes spatial homogeneity and isotropy by inserting the FLRW metric into equation~(\ref{gen_fieldeq}), the field equations reduce to the Friedmann and Raychaudhuri equations:
\begin{equation} \label{alk}
\begin{split}
3F\left(H^2 + \frac{k}{a^2}\right) &= \kappa\rho + \frac{1}{2}\left(RF - f\right) - 3H\dot{F}, \\
3H^2 + 3\dot{H} &= -\frac{1}{2F} \left( \kappa(\rho + 3P) + f - F R + 3 H F' \dot{R} + 3 F'' \dot{R}^2 + 3 F' \ddot{R} \right),
\end{split}
\end{equation}
where \( F = \frac{df}{dR} \), \(F^{\prime}=\frac{dF}{dR}\), \(\kappa=8\pi G\) and \( k \) is the spatial curvature parameter (\( k = 0, \pm1 \)). The quantities \( \rho \) and \( P \) denote the energy density and pressure, respectively, of the cosmic matter content, which is typically modeled as a perfect fluid. The pressure and energy density are related by an equation of state of the form \( P = w\rho \). The Friedmann equation in (\ref{alk}) can be recast in the form
\begin{equation}\label{afgo}
\left(3H+\frac{3}{2}\frac{\dot{F}}{F}\right)^2+\frac{3}{2}\left(\frac{f}{F}+\frac{6k}{a^2}\right)=\frac{3\rho}{F}+\frac{3R}{2}+\left(\frac{3}{2}\frac{\dot{F}}{F}\right)^2.
\end{equation}
To construct a compact phase space, one has to define the normalization factor
\begin{equation}\label{dee}
D^2 = \left(3H+\frac{3}{2}\frac{\dot{F}}{F}\right)^2 + \frac{3}{2}\left(\frac{f}{F} + \frac{6k}{a^2}\right),
\end{equation}
and introduce the dimensionless dynamical variables
\begin{equation}\label{algo}
\bar{x} = \frac{3}{2} \frac{\dot{F}}{F} \frac{1}{D}, \quad
\bar{y} = \frac{3}{2} \frac{R}{D^2}, \quad
\bar{z} = \frac{3}{2} \frac{f}{F} \frac{1}{D^2}, \quad
\bar{Q} = \frac{3H}{D}, \quad
\bar{\Omega} = \frac{3\rho}{F} \frac{1}{D^2}, \quad
\bar{K} = \frac{9k}{a^2} \frac{1}{D^2}.
\end{equation}
Expressing equation (\ref{afgo}) in terms of these variables results in the constraint:
\begin{equation}\label{abn}
\bar{\Omega} + \bar{y} + \bar{x}^2 = 1,
\end{equation}
while equation (\ref{dee}) becomes:
\begin{equation}\label{aon}
\left(\bar{Q} + \bar{x}\right)^2 + \bar{z} + \bar{K} = 1.
\end{equation}
The constraints $k \geq 0$, $R \geq 0$, $f \geq 0$, and $F > 0$ that naturally arise from a thorough study of $f(R)$ theories, when combined with (\ref{abn}) and (\ref{aon}), define a compact phase space with bounded dynamical variables:
\begin{align}
-1 &\leq \bar{x} \leq 1, \nonumber \\
-2 &\leq \bar{Q} \leq 2, \nonumber \\
0 &\leq \bar{y},\ \bar{z},\ \bar{\Omega} \leq 1. \label{bounds}
\end{align}
$H = 0$ corresponds to $\bar{Q} = 0$, which is permitted in this formulation. This compact phase space is especially well-suited to study cosmological bounces and non-singular scenarios. The dynamical equations associated with these variables (\ref{algo}) are a set of first order differential equations \cite{chakraborty2024model}:
\begin{align} 
\bar{x}' &= \frac{1}{6} \left[ 2 \left( -1 + 2 \bar{K} + \bar{x} \bar{Q} - 2 \bar{x}^2 \bar{K} + 2 \bar{Q}^2 - 2 \bar{x}^2 \bar{Q}^2 - \bar{x}^3 \bar{Q} + \bar{x}^4 \right) - \bar{\Omega} (1 + 3 w) \right. \nonumber \\
&\quad \left. + 3 \bar{x} \bar{\Omega} (1 + w) (\bar{Q} + \bar{x}) - \bar{x} \bar{Q}^3 (j - q - 2) + 2 \bar{x} \bar{K} \bar{Q} \right], \label{awq} \\
\bar{\Omega}' &= \frac{\bar{\Omega}}{3} \left[ 3 (\bar{x} \bar{\Omega} + \bar{Q} \bar{\Omega} - \bar{Q}) (1 + w) - \bar{Q}^3 (j - q - 2) + 2 \bar{K} \bar{Q} - 4 \bar{x} \bar{Q}^2 - 4 \bar{x} \bar{K} \right. \nonumber \\
&\quad \left. - 2 \bar{x}^2 \bar{Q} + 2 \bar{x}^3 \right], \label{aqr} \\
\bar{Q}' &= \frac{1}{6} \left[ 3 \bar{Q} \bar{\Omega} (1 + w) (\bar{Q} + \bar{x}) - 2 \bar{K} - 4 \bar{Q}^2 + 2 \bar{x} \bar{Q} - \bar{Q}^4 (j - q - 2) + 2 \bar{K} \bar{Q}^2 \right. \nonumber \\
&\quad \left. + 2 (1 - \bar{\Omega} - \bar{x}^2) - 4 \bar{x} \bar{Q} \bar{K} - 2 \bar{x}^2 \bar{Q}^2 + 2 \bar{x}^3 \bar{Q} - 4 \bar{x} \bar{Q}^3 \right], \label{afg} \\
\bar{K}' &= \frac{\bar{K}}{6} \left[ 4 \bar{x} - 4 \bar{Q} - 8 \bar{x} \bar{K} - 8 \bar{x} \bar{Q}^2 - 4 \bar{x}^2 \bar{Q} + 4 \bar{x}^3 + 6 \bar{\Omega} (1 + w) (\bar{Q} + \bar{x}) \right. \nonumber \\
&\quad \left. - 2 \bar{Q}^3 (j - q - 2) + 4 \bar{K} \bar{Q} \right]\,, \label{asx}
\end{align}
where $\bar{y}$ was eliminated using (\ref{abn}). The prime denotes differentiation of the dynamical variables \eqref{algo} with respect to the phase space time variable $\bar{\tau}$, defined by $d\bar{\tau} = D \, dt$, where $D > 0$ from the normalization condition (\ref{dee}). This formulation follows standard dynamical systems methodology \citep{bahamonde2018dynamical,Boehmer:2014vea,wainwright1989dynamical}, where physical states in phase space evolve via first-order differential equations. Here, the state is characterized by four compact variables \eqref{algo} (without $\bar{y}$), governed by the autonomous equations \eqref{awq}--\eqref{asx}, forming a four-dimensional system. In contrast, the non-compact formulation of $f(R)$ gravity \citep{chakraborty2024model} uses variables $(x,y,z,\Omega, K)$, related to the compact variables by
\begin{equation}\label{nonc}
x = \frac{2 \bar{x}}{\bar{Q}}, \quad
y = \frac{\bar{y}}{\bar{Q}^2}, \quad
z = \frac{\bar{z}}{\bar{Q}^2}, \quad
K = \frac{\bar{K}}{\bar{Q}^2}, \quad
\Omega = \frac{\bar{\Omega}}{\bar{Q}^2},
\end{equation}
which become undefined at $\bar{Q} = 0$ ($H = 0$).
Non-compact dynamical systems in $f(R)$ gravity fail to capture cosmological evolutions where $H = 0$, a critical feature of bouncing cosmologies. This motivated the development of compact dynamical systems \cite{goliath1999homogeneous}. The compact formulation addresses this limitation, as evident in \eqref{algo}, where $\bar{Q}=H=0$ is permitted. Notably, the dynamical equations \eqref{awq}--\eqref{asx} depend on cosmographic parameters $q$ and $j$ \eqref{acv} through the term $\bar{Q}^3 (j - q - 2)$, rather than the specific $f(R)$ function. For the power-law model \eqref{laqq}, this term is:
\begin{equation}\label{alls}
\bar{Q}^3 (j - q - 2) = -4 \bar{Q}^3 (q + 1) + \frac{2}{\beta^2} \bar{Q}^3,
\end{equation}
where $j - q - 2 = -4 (q + 1) + \frac{2}{\beta^2}$. Substituting
\begin{equation}\label{mal}
q + 1 = 2 - \frac{1 - \bar{\Omega} - \bar{x} - \bar{K}}{\bar{Q}^2}
\end{equation}
into \eqref{alls} yields:
\begin{equation}\label{azo}
\bar{Q}^3 (j - q - 2) = \bar{Q}^3 \left( \frac{2}{\beta} - 8 \right) - 4 \bar{Q} (1 - \bar{\Omega} - \bar{x} - \bar{K}).
\end{equation}
Obtained in \cite{chakraborty2024model}, equation \eqref{mal} links the cosmographic parameter $q$ to the dynamical variables \eqref{algo}, enabling the determination of whether the Friedmann-Lemaître-Robertson-Walker (FLRW) spacetime is accelerating or decelerating in $f(R)$ gravity for a given tuple $(\bar{x}, \bar{\Omega}, \bar{Q}, \bar{K})$. Substituting \eqref{azo} into the dynamical equations \eqref{awq}--\eqref{asx} results in the following autonomous system:
\begin{align}
\bar{x}^{\prime} &= \frac{1}{6} \Big[ 2 \big( -1 + 2 \bar{K} + \bar{x} \bar{Q} - 2 \bar{x}^2 \bar{K} + 2 \bar{Q}^2 - 2 \bar{x}^2 \bar{Q}^2 - \bar{x}^3 \bar{Q} + \bar{x}^4 \big) - \bar{\Omega} (1 + 3 w) \notag \\
&\quad + 3 \bar{x} \bar{\Omega} (1 + w) (\bar{Q} + \bar{x}) - \bar{x} \bar{Q}^3 \left( \frac{2}{\beta} - 8 \right) + 4 \bar{x} \bar{Q} (1 - \bar{\Omega} - \bar{x} - \bar{K}) + 2 \bar{x} \bar{K} \bar{Q} \Big], \label{comp1} \\
\bar{\Omega}^{\prime} &= \frac{\bar{\Omega}}{3} \Big[ 3 \big( \bar{x} \bar{\Omega} + \bar{Q} \bar{\Omega} - \bar{Q} \big) (1 + w) - \bar{Q}^3 \left( \frac{2}{\beta} - 8 \right) + 4 \bar{Q} (1 - \bar{\Omega} - \bar{x} - \bar{K}) \notag \\
&\quad + 2 \bar{K} \bar{Q} - 4 \bar{x} \bar{Q}^2 - 4 \bar{x} \bar{K} - 2 \bar{x}^2 \bar{Q} + 2 \bar{x}^3 \Big], \label{comp2} \\
\bar{Q}^{\prime} &= \frac{1}{6} \Big[ 3 \bar{Q} \bar{\Omega} (1 + w) (\bar{Q} + \bar{x}) - 2 \bar{K} - 4 \bar{Q}^2 + 2 \bar{x} \bar{Q} - \bar{Q}^4 \left( \frac{2}{\beta} - 8 \right) + 4 \bar{Q}^2 (1 - \bar{\Omega} - \bar{x} - \bar{K}) \notag \\
&\quad + 2 \bar{K} \bar{Q}^2 + 2 (1 - \bar{\Omega} - \bar{x}^2) - 4 \bar{x} \bar{Q} \bar{K} - 2 \bar{x}^2 \bar{Q}^2 + 2 \bar{x}^3 \bar{Q} - 4 \bar{x} \bar{Q}^3 \Big], \label{comp3} \\
\bar{K}^{\prime} &= \frac{\bar{K}}{6} \Big[ 4 \bar{x} - 4 \bar{Q} - 8 \bar{x} \bar{K} - 8 \bar{x} \bar{Q}^2 - 4 \bar{x}^2 \bar{Q} + 4 \bar{x}^3 + 6 \bar{\Omega} (1 + w) (\bar{Q} + \bar{x}) \notag \\
&\quad - 2 \bar{Q}^3 \left( \frac{2}{\beta} - 8 \right) + 8 \bar{Q} (1 - \bar{\Omega} - \bar{x} - \bar{K}) + 4 \bar{K} \bar{Q} \Big]. \label{comp4}
\end{align}
Next, we analyze the fixed points of \eqref{comp1}–\eqref{comp4} for $\beta > 1$ (accelerating regime) and $\beta < \frac{1}{3}$ (ekpyrotic regime), expressed as tuples $(\bar{x}, \bar{\Omega}, \bar{Q}, \bar{K})$, to compute $q$ via \eqref{mal} and characterize the cosmological evolution.\\

 Note that, in deriving the dynamical system equations \eqref{comp1}–\eqref{comp4}, we employ a power-law scale factor as an ansatz to simplify the analysis of the ekpyrotic and accelerating regimes. We clarify that this form is a local approximation, used to facilitate the study of phase space dynamics in specific cosmological epochs, and is not assumed to describe the entire evolution, particularly at the bounce. The bounce, defined by $(H = 0)$ at a finite time, is captured by phase space trajectories crossing $(\bar Q = 0)$, enabled by the modified dynamics of $f(R)$ gravity, which will permits violations of the null energy condition as discussed in section \ref{sec:weff}. This ensures our dynamical systems approach will consistently model the bounce.
\section{Effective Equation of State in \texorpdfstring{$f(R)$}{f(R)} Gravity}
\label{sec:weff}

Before proceeding with the analysis of fixed points and cosmological bounces, it is crucial to discuss the effective equation of state parameter \( w_{\text{eff}} \) in \( f(R) \) gravity, as it provides insight into the cosmological dynamics and the nature of the bounce. In \( f(R) \) gravity, the modified field equations (\ref{alk}) can be recast to resemble the standard Friedmann and Raychaudhuri equations of general relativity, but with an effective energy-momentum tensor that encapsulates the contributions of the modified gravity terms. This effective energy-momentum tensor can be characterized by an effective equation of state parameter \( w_{\text{eff}} \), defined as:
\begin{equation}
w_{\text{eff}} = \frac{P_{\text{eff}}}{\rho_{\text{eff}}},
\end{equation}
where \( \rho_{\text{eff}} \) and \( P_{\text{eff}} \) are the effective energy density and pressure, respectively, derived from the \( f(R) \) contributions and the matter content. For a flat FLRW spacetime (\( k = 0 \)), the effective energy density and pressure can be expressed as \cite{DeFelice:2010aj,Sotiriou:2008rp}:
\begin{align}\label{frg}
\rho_{\text{eff}} &= \frac{1}{F} \left[ \kappa \rho + \frac{1}{2} (R F - f) - 3 H \dot{F} \right], \\
P_{\text{eff}} &= \frac{1}{F} \left[ \kappa P + \ddot{F} + 2 H \dot{F} - \frac{1}{2} (R F - f) \right],
\end{align}
where 
\( \kappa = 8\pi G \), and \( \rho \) and \( P \) are the matter density and pressure related by \( P = w \rho \). The effective equation of state parameter \( w_{\text{eff}} \) governs the overall dynamics of the universe, including whether it undergoes accelerated expansion (\( w_{\text{eff}} < -\frac{1}{3} \)), deceleration (\( w_{\text{eff}} > -\frac{1}{3} \)), or a bounce scenario.\\

In the context of our dynamical systems approach, \( w_{\text{eff}} \) can be related to the cosmographic parameters, particularly the deceleration parameter \( q \), via:
\begin{equation}\label{eff}
w_{\text{eff}} = \frac{1}{3} (2q - 1).
\end{equation}
This relation arises because the effective Friedmann equation in \(f(R)\) gravity mimics the form \(\dot{H} = -\frac{\kappa^2}{2} \left(\rho_{\text{eff}} + P_{\text{eff}}\right)\), which is valid in spatially flat models. 
For a bounce to occur, the universe must transition from contraction (\( H < 0 \)) to expansion (\( H > 0 \)) at \( H = 0 \), often requiring \( w_{\text{eff}} < -1 \) (phantom-like behavior) to violate the null energy condition, which is a hallmark of non-singular bounces in modified gravity theories \citep{bamba2014bounce, buchbinder2007new}.\\

The matter equation of state parameter \( w \) plays a critical role in determining \( w_{\text{eff}} \). In standard cosmology, typical values of \( w \) correspond to different types of matter: \( w = 0 \) for dust (pressureless matter, relevant for the matter-dominated epoch), \( w = \frac{1}{3} \) for radiation (relevant for the radiation-dominated epoch), and \( w = -1 \) for a cosmological constant or dark energy. In \( f(R) \) gravity, the modified gravitational dynamics can amplify or alter the effective behavior of the matter content, leading to a \( w_{\text{eff}} \) that differs from \( w \). For instance, even with \( w = 0 \) or \( w = \frac{1}{3} \), the \( f(R) \) terms can induce \( w_{\text{eff}} < -1 \), facilitating a bounce.

For the analysis in this paper, we focus on two key values of \( w \):

\begin{itemize}
    \item \( w = 0 \) (Dust): This corresponds to pressureless matter, typical of the matter-dominated epoch in standard cosmology. In \( f(R) \) gravity, dust-dominated universes can exhibit non-singular bounces if the modified gravity terms produce a sufficiently negative \( w_{\text{eff}} \). This case is analyzed in Section \ref{dust} to explore bounces in both ekpyrotic (\( \beta < \frac{1}{3} \)) and accelerating (\( \beta > 1 \)) regimes.
    \item \( w = \frac{1}{3} \) (Radiation): This corresponds to relativistic matter, relevant for the radiation-dominated epoch. In \( f(R) \) gravity, radiation-dominated scenarios can also lead to bounces, particularly in the ekpyrotic regime, where the interplay between matter and modified gravity effects can drive a transition through \( H = 0 \). This case is also analyzed in Section \ref{rad}.
\end{itemize}

Additionally, it is worth considering \( w = -1 \) (scalar field or dark energy-like matter), as it naturally leads to accelerated expansion and could enhance bounce scenarios in the accelerating regime (\( \beta > 1 \)). However, since our focus is on early-time cosmology and ekpyrotic scenarios, we prioritize \( w = 0 \) and \( w = \frac{1}{3} \), as these are more relevant to the early universe's matter and radiation-dominated phases.\\ 

By examining \( w_{\text{eff}} \) through the dynamical variables (\( \bar{x}, \bar{\Omega}, \bar{Q}, \bar{K} \)) and their relation to \( q \), we can better understand the conditions under which \( f(R) \) gravity supports non-singular bounces. The compact dynamical system framework, introduced in this section, allows us to track the evolution of \( w_{\text{eff}} \) across the phase space, particularly at fixed points where \( \bar{Q} = 0 \) (corresponding to \( H = 0 \)), which are critical for identifying bounce solutions.

\noindent
\subsection*{Case 1: Pressureless Matter (Dust), \(w=0\):}\label{dust}
In this subsection we investigate cosmological bounces in the dust-dominated epoch, characterized by the equation of state parameter $w = 0$, using the autonomous dynamical system (\ref{comp1})--(\ref{comp4}) in $f(R)$ gravity.\\

{\bf Fixed Points and Cosmology:}\label{sec:bounce} 
The fixed points of the dynamical system given by equations \eqref{comp1}--\eqref{comp4} in the case when the equation of state parameter $w=0$ (dust) and $\beta>1$ (accelerating solution), are found to be
\begin{flalign}
&\begin{aligned}
E &= (\bar{x}, \bar{\Omega}, \bar{Q}, \bar{K}) = (\frac{1}{10},\frac{93}{250}, -\frac{2}{5}, \frac{289}{500}) \\
F &= (\bar{x}, \bar{\Omega}, \bar{Q}, \bar{K}) = (0, \frac{2}{5}, 0, \frac{3}{5}) \;.
\end{aligned} \label{eq:fixed_points}
\end{flalign}
Following \cite{chakraborty2024model}, the cosmology associated with these fixed points is determined using equation (\ref{mal}). For fixed point $E$, the cosmology is found to be
\begin{equation}\label{asoh}
q = 1 - \frac{1 - \frac{93}{250}- \frac{1}{10} - \frac{289}{500}}{(-\frac{2}{5})^2}> 0\,,
\end{equation}
with $H=\bar{Q}=-\frac{2}{5}<0$ which describes the decelerated contracting phase before the big bang/bounce, whereas the cosmology associated with fixed point $F$ is indeterminate.
We will address this in the next section. The cosmologies corresponding to these fixed points is summarized in Table \ref{table1}.
\begin{table}
\centering
\begin{tabular}{ |p{3cm}|p{4cm}|p{3cm}|  }
\hline
\multicolumn{3}{|c|}{Cosmology of the fixed points when $w=0$, $\beta>1$ and $\beta<\frac{1}{3}$} \\
\hline
\centering{\textbf{Fixed point}} &\centering{\textbf{Coordinates}}\\ ($\bar{x},\bar{\Omega},\bar{Q},\bar{K})$ & \textbf{Cosmology} \\[10pt]
\hline
\centering{E}&\centering{$(\frac{1}{10},\frac{93}{250},-\frac{2}{5},\frac{289}{500})$}&$q>0$\\[10pt]
\hline
\centering{G} &\centering{$(1,-\frac{12}{5},-2,-\frac{8}{5})$}& $q=0$\\[10pt]
\hline
\centering{F}  & \centering{$(0,\frac{2}{5},0,\frac{3}{5})$} & Indeterminate\\[10pt]
\hline
\end{tabular}
\caption{The cosmology associated with each fixed point for $w=0$, with $\beta>1$ and $\beta<\tfrac{1}{3}$. The bounce is described by the fixed point $F$, for which the deceleration parameter $q$ is indeterminate. The remaining fixed points, $E$ and $G$, correspond to the decelerated contraction phase preceding the bounce and to cosmological solutions that are not relevant for the present analysis, respectively.}

\label{table1}
\end{table}
\begin{table}[h!]
\centering
\renewcommand{\arraystretch}{1.5}
\begin{tabular}{|c|p{7.8cm}|c|}
\hline
\multicolumn{3}{|c|}{Cosmology of the fixed points when $w=\frac{1}{3}$, $\beta>1$ and $\beta<\frac{1}{3}$} \\
\hline
\textbf{Fixed point} & \centering\textbf{Coordinates}\\ $(\bar{x},\bar{\Omega},\bar{Q},\bar{K})$ & \textbf{Cosmology} \\
\hline
A & \multicolumn{1}{c|}{\small{$\left(\frac{1}{3},\ \frac{1}{6},\ -\frac{1}{3},\ \frac{11}{18}\right)$}} & $q>0$ \\
\hline
B & \multicolumn{1}{c|}{\small{$\left(-\frac{1}{19},\ 1.70\times 10^{-1},\ \frac{1}{19},\ 6.25\times 10^{-1}\right)$}} & $q <-1$ \\
\hline
C & \multicolumn{1}{c|}{\small{$\left(\frac{3}{8},\ 0,\ 0,\ \frac{5}{8}\right)$}} & Indeterminate \\
\hline
D & \footnotesize{$\left(-5.81\times 10^{-2},\ 3.69\times 10^{-1},\ 5.81\times 10^{-2},\ 6.25\times 10^{-1}\right)$} & $q <-1$ \\
\hline
V & \footnotesize{$\left(8.20\times 10^{-1},\ 8.86\times 10^{-1},\ -8.20\times 10^{-1},\ 5.41\times 10^{-1}\right)$} & $q>0$ \\
\hline
\end{tabular}
\caption{The cosmology associated with each fixed point for $w=\tfrac{1}{3}$, with $\beta>1$ and $\beta<\tfrac{1}{3}$. The bounce is described by the fixed point $C$, for which the deceleration parameter $q$ is indeterminate. In the ekpyrotic regime ($\beta < \tfrac{1}{3}$), fixed point $V$ corresponds to the pre-bounce contracting phase, 
while $D$ corresponds to post-bounce expansion. In the accelerating regime ($\beta > 1$), the pre-bounce contraction 
is represented by fixed point $A$, and post-bounce expansion by fixed point $B$.}
\label{table1a}
\end{table}
The fixed points of the dynamical system \eqref{comp1}--\eqref{comp4} when $w=0$ and $\beta<\frac{1}{3}$ (ekpyrosis) is $F$ along with
\begin{equation}
 G=(1,-\frac{12}{5},-2,-\frac{289}{500})\,.
\end{equation}
When we insert $G$ into (\ref{mal}) we find that
\begin{equation}\label{azc}
q=0\,.
\end{equation}
The stability of these fixed points when $\beta>1$ ($\beta<\frac{1}{3}$) and $w=0$ is presented in Table \ref{table3}.\\
\begin{table}[h!]
\centering
\renewcommand{\arraystretch}{1.5}
\begin{tabular}{ |p{3cm}|p{6cm}|p{3cm}| }
\hline
\multicolumn{3}{|c|}{Stability of the fixed points when $w=0$, $\beta>1$ and $\beta<\frac{1}{3}$} \\
\hline
\textbf{Fixed point} & \centering\textbf{Sign of eigenvalues} \\ $(\lambda_1,\lambda_2,\lambda_3,\lambda_4)$ & \textbf{Stability} \\
\hline
\centering{E} & \centering{($-$, $z$, $z^*$, $-$)} & Saddle \\
\hline
\centering{G} & \centering{($+$, $+$, $-$, $+$)} & Saddle \\
\hline
\centering{F} & \centering{($+$, $z$, $z^*$, $-$)} & Saddle \\
\hline
\end{tabular}
\caption{The stability of the fixed points for the case $\beta > 1$, $\beta<\frac{1}{3}$ and $w = 0$ is summarized. Complex eigenvalues are denoted by $z = a + ib$ and $z^* = a - ib$, where $a$ and $b$ vary per fixed point (values in Appendix).}
\label{table3}
\end{table}

\begin{table}[h!]
\centering
\renewcommand{\arraystretch}{1.5}
\begin{tabular}{|c|>{\centering\arraybackslash}p{6.5cm}|>{\centering\arraybackslash}p{3cm}|}
\hline
\multicolumn{3}{|c|}{Stability of the fixed points when $w=\frac{1}{3}$, $\beta>1$ and $\beta<\frac{1}{3}$} \\
\hline
\textbf{Fixed point} & \centering{\textbf{Sign of eigenvalues}}\\ $(\lambda_1,\lambda_2,\lambda_3,\lambda_4)$ & \textbf{Stability} \\
\hline
A & ($-$, $z$, $z^*$, $+$) & Saddle \\
\hline
B & ($-$, $z$, $z^*$, $+$) & Saddle \\
\hline
C & ($-$, $+$, $z$, $z^*$) & Saddle \\
\hline
D & ($+$, $-$, $z$, $z^*$) & Saddle \\
\hline
V & ($+$, $+$, $+$, $+$) & Unstable \\
\hline
\end{tabular}
\caption{The stability of the fixed points for the case $\beta > 1$, $\beta < \frac{1}{3}$, and $w = \frac{1}{3}$ is summarized. Complex eigenvalues of the Jacobian matrix are denoted by $z = a + ib$ and their complex conjugates by $z^* = a - ib$, where $a$ and $b$ are the real and imaginary parts, respectively. The specific values of $a$ and $b$ for fixed point $F$ and $C$ are provided in the appendix. Note that these values vary between fixed points.} 
\label{table2c}
\end{table}


\noindent
Note that the deceleration parameter $q$ can be expressed as
\begin{equation}\label{ghy}
q=-\frac{\ddot{a}a}{\dot{a}^2}.
\end{equation}
When we set $q$ to zero in the above equation (\ref{ghy}) we find that
\begin{equation}\label{asg}
a(t)=0 \,\,\,\text{or}\,\, a(t) \propto t.
\end{equation}
The solution \( a(t) \propto t \), corresponding to a coasting cosmology with constant expansion rate, implies \( \beta = 1 \). This specific case is excluded from the class of power-law models considered in our analysis, which are restricted to the regimes \( \beta < \frac{1}{3} \) (ekpyrotic) and \( \beta > 1 \) (accelerating expansion). This leaves \( a(t) = 0 \) as the only viable solution, corresponding to the Big Bang singularity. The saddle fixed point with \( q = 0 \) (fixed point \( G \)) act as a repeller on one side and an attractor on the other, with the Big Bang singularity positioned between them. This configuration, in principle, suggests that it is possible to bypass the Big Bang singularity in \( f(R) \) theories of gravity. However, the bypass is not a conventional bounce. Phase space trajectories converge at the singular point and, rather than terminating, continue through it while maintaining \( a(t) = 0 \). The condition required to violate the convergence criterion consistent with bouncing models in \( f(R) \simeq R^n \) theories discussed in \cite{das2024cosmological,burger2018towards,haque2017consistent} was found to be
\begin{equation}\label{cons}
\frac{3}{n}\frac{\ddot{a}}{a}+\left(\frac{3}{n}-3\right)\frac{\dot{a}^2}{a^2}\geq 0.
\end{equation}
When we express equation (\ref{cons}) in terms of the cosmographic parameters defined in (\ref{acv}), we obtain the following constraint on the deceleration parameter $q$:
\begin{equation}\label{qu}
q \leq 1 - n\,,
\end{equation}
which holds for models with $n > 2$ (note that models with $n < 2$ do not violate the convergence condition). The value $q = 0$ does not satisfy the constraint in (\ref{qu}). This supports our earlier claim that the $q = 0$ solution does not correspond to a bounce, but rather represents an unconventional way of bypassing the singularity, as previously discussed. In the next subsection, we will particularly focus on the indeterminate cosmology, which gives rise to interesting physics for cosmological bounces.\\
\begin{figure}[h]
\begin{subfigure} {0.5\textwidth}
\includegraphics[width=1\linewidth, height=6cm]{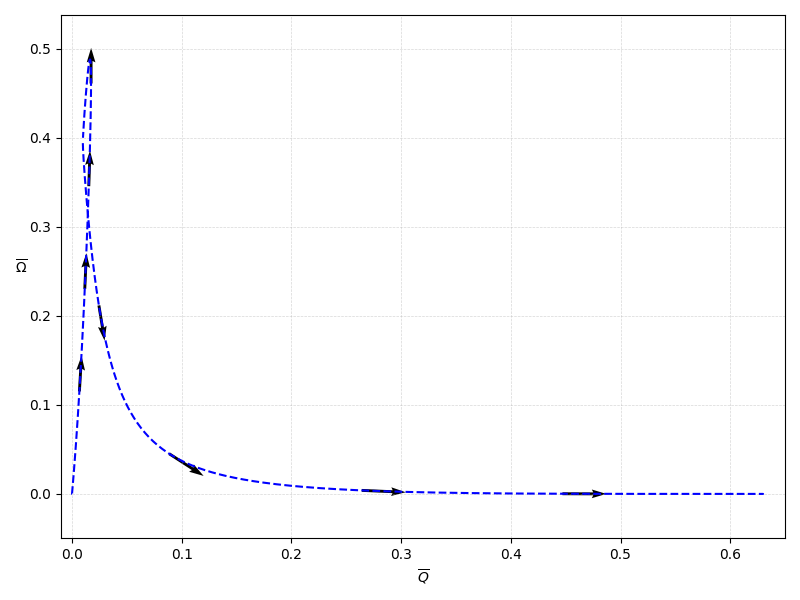}
\end{subfigure}
\hspace{1em}
\begin{subfigure}{0.5\textwidth}
\includegraphics[width=1\linewidth, height=6cm]{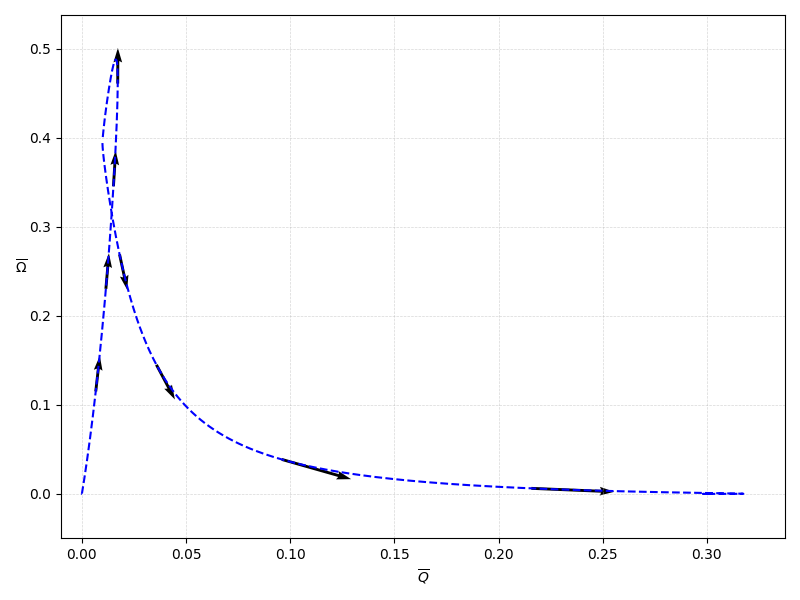}
\end{subfigure}
\caption{
Phase space trajectories perturbed from the fixed point \( F = (\bar{\Omega}, \bar{Q}) = (0.4, 0) \) are shown in the \( \bar{\Omega} \)--\( \bar{Q} \) plane. \textbf{Left plot} illustrates the ekpyrotic regime (\( \beta < \tfrac{1}{3} \), \( w = 0 \)). The trajectories evolve from the minimum (\( \bar{Q} = 0 \)) to expansion (\( \bar{Q} > 0 \)), indicating a cosmological bounce. \textbf{Right plot} illustrates the accelerating regime (\( \beta > 1 \), \( w = 0 \)). A bounce is again present, with the trajectories exhibiting pronounced post-bounce expansion characterized by \( \bar{Q} > 0 \).
}
\label{fig1}
\end{figure}
\begin{figure}[h]
\begin{subfigure} {0.5\textwidth}
\includegraphics[width=1\linewidth, height=6cm]{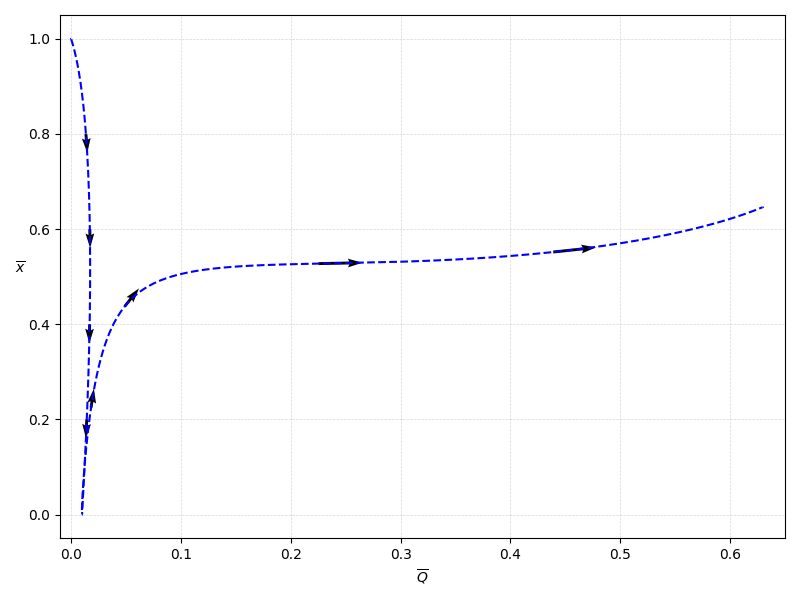}
\end{subfigure}
\hspace{1em}
\begin{subfigure}{0.5\textwidth}
\includegraphics[width=1\linewidth, height=6cm]{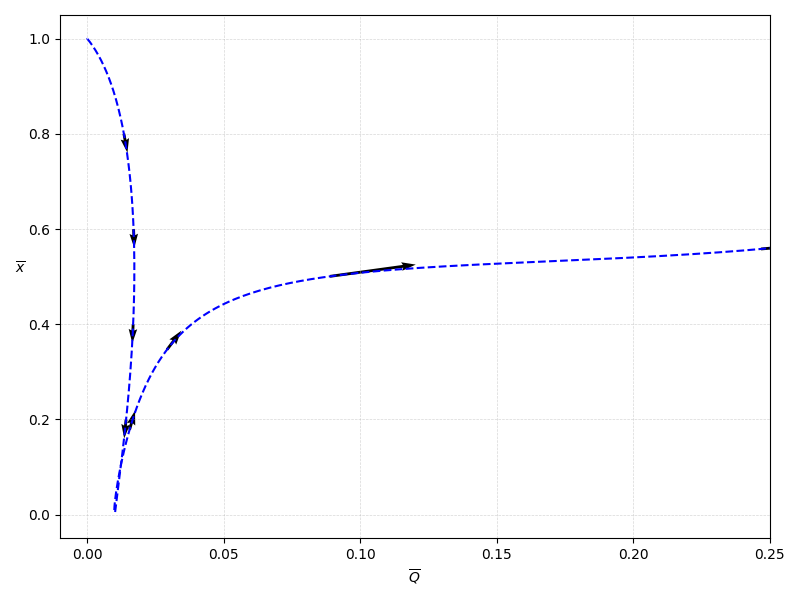}
\end{subfigure}
\caption{Phase space trajectories perturbed from the fixed point $F$ ($(\bar{x}, \bar{Q}) = (0, 0)$) are shown in the $\bar{x}$--$\bar{Q}$ plane. \textbf{Left plot} illustrates the ekpyrotic regime ($\beta < \tfrac{1}{3}$, $w = 0$). The trajectories cross the $\bar{Q} = 0$ axis, indicating a bounce from the minimum value of $\bar{Q}$ to expansion. \textbf{Right plot} illustrates the accelerating regime ($\beta > 1$, $w = 0$). The bounce is again evident, with trajectories extending into the expanding phase where $\bar{Q} > 0$.}
\label{fig2}
\end{figure}
\begin{figure}[h]
\begin{subfigure} {0.5\textwidth}
\includegraphics[width=1\linewidth, height=6cm]{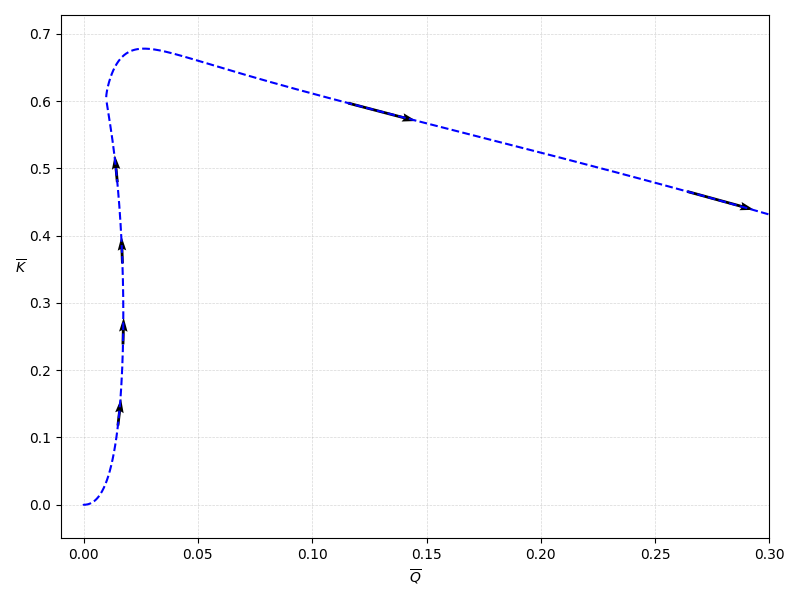}
\end{subfigure}
\hspace{1em}
\begin{subfigure}{0.5\textwidth}
\includegraphics[width=1\linewidth, height=6cm]{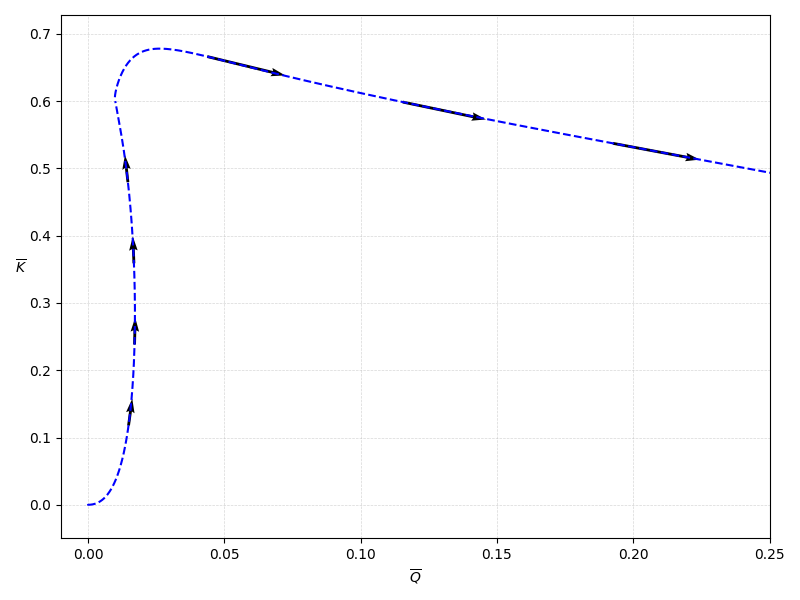}
\end{subfigure}
\caption{Phase space trajectories perturbed from the fixed point $F$ ($(\bar{K}, \bar{Q}) = (0.6, 0)$) are shown in the $\bar{K}$--$\bar{Q}$ plane. \textbf{Left plot} illustrates the ekpyrotic regime ($\beta < \tfrac{1}{3}$, $w = 0$). The trajectories exhibit a bounce, transitioning through $\bar{Q} = 0$ from contraction to expansion. \textbf{Right plot} illustrates the accelerating regime ($\beta > 1$, $w = 0$), the bounce is again confirmed, with trajectories displaying strong post-bounce expansion characterized by $\bar{Q} > 0$.}
\label{fig3}
\end{figure}

\noindent
\textbf{Indeterminate Cosmology and Bounce:}\label{sec:perturbation}
 The fixed points with an indeterminate cosmology all share the feature $\bar{Q} = 0$, which, according to equation (\ref{algo}), occurs either when $H = \frac{\dot{a}}{a} = 0$ or when $a(t) \propto \lvert t \rvert^{\beta}$ as $\lvert t \rvert \to \infty$. The latter scenario is not applicable in our case, as we are only interested in early-time solutions. The former case, $H = 0$, implies that the scale factor is finite and at a minimum. The key question, however, is whether it remains at this minimum indefinitely.\\
 
To address this question more precisely, we perturb the fixed point $F$ as follows:
\begin{equation}\label{assip}
F = \bar{F} + \delta F,
\end{equation}
where $\delta F$ represents a first-order perturbation around the fixed point $\bar{F}$.\\
 
Now the dynamical equations \eqref{comp1}--\eqref{comp4} can be put into the following matrix form 
\begin{equation} F^{\prime}=JF, \end{equation} where $F'=(\bar x',\bar \Omega', \bar Q', \bar K')$ is a column matrix and $J$ is the Jacobian matrix of the dynamical system. Inserting (\ref{assip}) into this dynamical equation results in 
\begin{equation}\label{asz}
\bar{F}^{\prime}+\delta{F}^{\prime} =J\left(\bar{F}+\delta F\right).
\end{equation}
The zeroth-order term in equation (\ref{asz}) yields $\bar{F}^{\prime} = J \bar{F}$, while the first-order equation takes the form:
\begin{equation}\label{pert}
\delta F^{\prime}=J\delta F.
\end{equation}
An immediate solution to the first order equation (\ref{pert}) is
\begin{equation}\label{acb}
\delta F=\sum_i c_iv_ie^{\lambda_i \bar \tau}\,,
\end{equation}
where $v_i$ are eigenvectors associated with the eigenvalues $\lambda_i$ of the Jacobian matrix $J_F$ evaluated at the fixed point $F$. The above solution (\ref{acb}) only holds if there are no repeated eigenvalues. Note also that $\delta F=(\delta\bar{x},\delta\bar{\Omega},\delta{\bar{Q}},\delta{\bar{K}})$ which implies that
\begin{equation}\label{ach}
\delta \bar{Q}=\sum_i c_iv_{i3}e^{\lambda_i\bar \tau}.
\end{equation}
The Jacobian $J_F$ has only one real and positive eigenvalue $\lambda_{F1}=1.15953$ (see appendix for the details). 
To analyze the post-bounce expanding phase we consider the perturbation is along $\lambda_{F1}$ and use the associated eigenvector whose third component is $v_{13}=0.194021$. Setting $c_i=0$ for $i \neq 1$ we obtain,
\begin{equation}\label{eq:3.29}
\delta \bar{Q}\simeq0.194021 e^{1.15953 \bar \tau}>0.
\end{equation}
Since $\bar Q \simeq \frac{3 H}{D}$ and $\bar F=0$ at the fixed point $F$ this implies
\begin{equation}\label{negv1}
H\simeq \delta\bar{Q}>0.
\end{equation}
Therefore, we see that the Hubble parameter $H$ grows--it does not stay at a minimum forever. Our analysis here is based on the phase space time variable $\bar{\tau}$, rather than cosmic time $t$. However, due to the monotonic relationship $d\bar{\tau} = D dt$, where $D > 0$, our arguments remain valid for cosmic time as well. To explicitly confirm that the scale factor reaches a finite minimum at a finite time $\bar \tau$ (hence $t$), we substitute equation (\ref{ach}) into (\ref{negv1}) as follows:
\begin{equation}\label{eq31}
\dot H \simeq \frac{v_{13}}{3} e^{\lambda_1 \bar \tau} (\dot D + \lambda_1 D^2).
\end{equation}
Since the quantity in the parenthesis is positive, $\dot H>0$ and consequently we have a bounce.

Expanding the perturbation sum along the eigenvector associated with the eigenvalue $\lambda_{F4} = -0.73004$ of the Jacobian matrix $J_F$, we obtain
\begin{equation}\label{alm}
H \simeq \delta \bar{Q} \simeq -4.27489 e^{-0.73004 \bar{\tau}} < 0.
\end{equation}
This indicates that the Hubble parameter $H$ is negative, corresponding to a contracting phase. The exponential term $e^{-0.73004 \bar{\tau}}$ grows for $\bar{\tau} < 0$, implying that this contraction occurs in the negative time direction, consistent with the pre-bounce phase of an ekpyrotic cosmology. In contrast, the perturbation along the positive eigenvalue $\lambda_{F1} = 1.15953$ drives expansion ($H > 0$), as shown in equation (\ref{eq:3.29}). The remaining eigenvalues of $J_F$ are complex and may indicate oscillatory behavior, though their physical significance in this context requires further investigation. \\

As an example, phase space trajectories, shown in Figures \ref{fig1}--\ref{fig3}, demonstrate that the fixed point $F$ corresponds to a cosmological bounce, where the scale factor contracts to a finite minimum value and subsequently expands.\\


\subsection*{Case 2: Radiation Dominated Epoch, \texorpdfstring{$w = \frac{1}{3}$}{w = 1/3}:}\label{rad}
We analyze the dynamics of the radiation-dominated epoch, characterized by the equation of state parameter $w = \frac{1}{3}$, using the autonomous system (\ref{comp1})--(\ref{comp4}) in the ekpyrotic ($\beta < \frac{1}{3}$) and accelerating ($\beta > 1$) regimes. Fixed points $V$, $D$, and $C$ (ekpyrotic) and $A$, $B$, and $C$ (accelerating), listed in Table \ref{table1a} \footnote{The large numerical values in the coordinates of fixed points \( V \), \( D \), and \( B \) in Table~\ref{table1a} 
arise due to the normalization factor \( D^2 = \left(3 H + \frac{3}{2} \frac{\dot{F}}{F}\right)^2 + \frac{3}{2} \left( \frac{f}{F} + \frac{6 k}{a^2} \right)\), which can amplify small differences in the dynamical variables. To ensure robustness, we recomputed the fixed points using a high-precision numerical solver 
and verified consistency within a relative error of \( 10^{-10} \). To improve readability, we approximate these coordinates in decimal form in Table~\ref{table1a}, with exact fractions provided in the appendix for reference.}, govern the cosmological evolution, with their stability detailed in Table \ref{table2c}. \\  


In the ekpyrotic regime ($\beta < \frac{1}{3}$), fixed point $V$, with $q >0$ and $\bar{Q} < 0$, is an unstable source (eigenvalues all positive, Table \ref{table2c}), initiating a contracting phase ($H \simeq \bar{Q} < 0$). Fixed point $D$, with $q<-1$ and $\bar{Q} > 0$, is a saddle point, representing post-bounce expansion. 
Fixed point $C$ has $\bar{Q} = 0$ (i.e., $H = 0$) and indeterminate $q$, marking the bounce. Its saddle nature 
facilitates the transition from contraction to expansion. Perturbing around $C$ as $C = \bar{C} + \delta C$, the perturbation in $\bar{Q}$ is $\delta \bar{Q} = \sum_i c_i v_{i3} e^{\lambda_i \bar \tau}$. For $\lambda_{C2}$, with eigenvector $\mathbf{v}_{C2}$
we obtain 
\begin{equation} \label{foc}
\delta \bar{Q} \simeq c_2 (0.8029) e^{0.17362 \bar \tau} > 0, \end{equation} 
(for $c_2 > 0$), indicating post-bounce expansion ($H > 0$). For $\lambda_{C1}$, with $\mathbf{v}_{C1}$ 
we find \begin{equation}\label{fac}
    \delta \bar{Q} \simeq c_1 (-0.9052) e^{-\frac{71}{256} \bar \tau} < 0,
    \end{equation}
    (for $c_1 > 0$), corresponding to pre-bounce contraction ($H < 0$). The vacuum state ($\bar{\Omega} = 0$) at $C$ implies that $f(R)$ gravity terms (\ref{frg}) dominate, as the matter density $\rho \propto a^{-4}$ is suppressed relative to modified gravity contributions.\\

In the accelerating regime ($\beta > 1$), fixed point $A$, with $q >0$ and $\bar{Q} = -\frac{1}{3} < 0$, is a saddle point, initiating contraction. Fixed point $B$, with $q <-1$ and $\bar{Q} > 0$, is a saddle point, with $\dot{H} = -(q + 1)H^2 > 0$, indicating strong post-bounce acceleration. Fixed point $C$, identical to the ekpyrotic case, facilitates the bounce. The perturbation analysis for $C$ applies identically, confirming the transition through $\bar{Q} = 0$. Unlike the ekpyrotic regime, where contraction begins from a source ($V$), here it starts from a saddle ($A$), reflecting a different dynamical origin but converging to the same bounce mechanism.\\

\begin{figure}[h]
\includegraphics[scale=0.7]{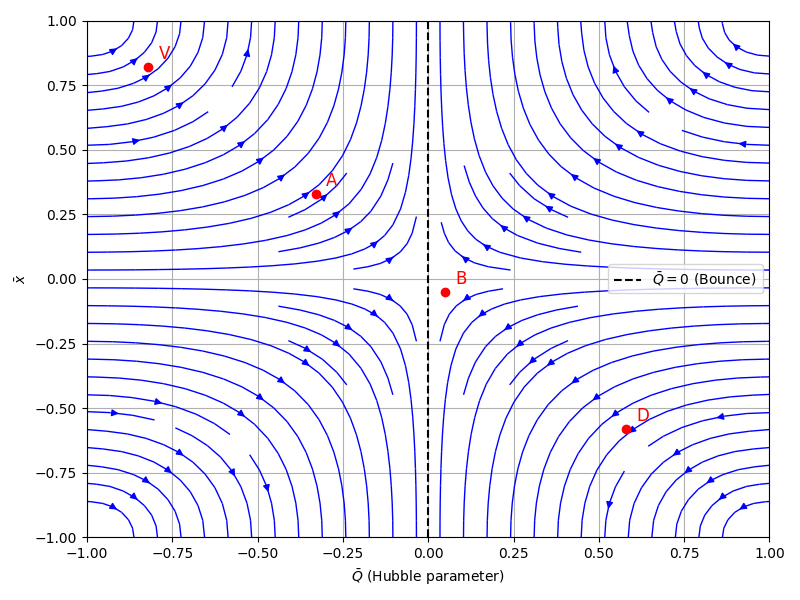}
\caption{Illustrative 2D phase portrait in the $(\bar{Q}, \bar{x})$ plane showing qualitative trajectories associated with a cosmological bounce. The vertical dashed line at $\bar{Q} = 0$ denotes the bounce surface where the Hubble parameter vanishes. Red points mark representative fixed points, with $V$ corresponding to a contracting source, $D$ to a post-bounce expanding saddle, and $C$ (not shown) expected near the bounce at $\bar{Q} = 0$. The flow pattern reflects the qualitative structure described in the analysis of Tables~\ref{table1a} and~\ref{table2c}.}
\label{illu}
\end{figure}

Phase space trajectories, illustrated in Figure \ref{illu}, show a qualitative flow from $V$ or $A$ (contraction) through $\bar{Q} = 0$ (bounce at $C$) to $D$ or $B$ (expansion). The saddle nature of the point C and the perturbation analysis confirm its role as the bounce point. \\

The effective equation of state $w_{\text{eff}} = \frac{1}{3}(2q - 1)$ yields $w_{\text{eff}} < -1$ at $D$ and $B$, supporting the violation of the null energy condition required for a non-singular bounce \citep{bamba2014bounce}. This analysis extends the findings for $w = 0$, demonstrating robust bounce solutions in the radiation-dominated epoch as well.

\section{Results}\label{sec:conclusion} 
In this paper, we explore cosmological bounces in FLRW spacetime within \( f(R) \) gravity, utilizing compact, model-independent dynamical systems characterized by power-law scale factors. 
Our analysis focuses on ekpyrotic (\( \beta < \frac{1}{3} \)) and accelerating (\( \beta > 1 \)) regimes, employing compact phase space variables \( (\bar{x}, \bar{\Omega}, \bar{Q}, \bar{K}) \), defined in (\ref{algo}), and autonomous differential equations (\ref{comp1})--(\ref{comp4}) to identify fixed points corresponding to non-singular bounces. A bounce occurs when the universe transitions from contraction (\( H < 0 \)) to expansion (\( H > 0 \)) at \( \bar{Q} = 0 \) (i.e., \( H = 0 \)), maintaining a finite scale factor. We consider two matter equation of state parameters, \( w = 0 \) (dust) and \( w = \frac{1}{3} \) (radiation), relevant to the early universe's matter and radiation-dominated epochs. The effective equation of state parameter \( w_{\text{eff}} = \frac{1}{3}(2q - 1) \), derived from equation (\ref{eff}), governs the dynamics, with \( w_{\text{eff}} < -1 \) facilitating bounces by violating the null energy condition \citep{bamba2014bounce}.\\

For the dust-dominated case (\( w = 0 \)), we identify fixed points \( E \), \( F \), and \( G \), as listed in Table \ref{table1}. Fixed point \( F \), located at \( (\bar{x}, \bar{\Omega}, \bar{Q}, \bar{K}) = (0, \frac{2}{5}, 0, \frac{3}{5}) \), is critical for the bounce, with \( \bar{Q} = 0 \) indicating \( H = 0 \). The deceleration parameter \( q \) at \( F \) is indeterminate due to division by \( \bar{Q}^2 = 0 \) in (\ref{mal}), but perturbation analysis clarifies the dynamics. Perturbing around \( F \) as \( F = \bar{F} + \delta F \), the Jacobian matrix \( J_F \) (\ref{JF}) yields real eigenvalues \( \lambda_{F1} = 1.15953 \), \( \lambda_{F4} = -0.73004 \), and a complex conjugate pair (Table \ref{table3}). The positive eigenvalue \( \lambda_{F1} \) drives post-bounce expansion, with the perturbation \( \delta \bar{Q} 
> 0 \) (\ref{eq:3.29}), implying \( H > 0 \). The negative eigenvalue \( \lambda_{F4} \) corresponds to pre-bounce contraction, with \( \delta \bar{Q} 
< 0 \) (\ref{alm}), indicating \( H < 0 \). Phase space trajectories, shown in Figures \ref{fig1}--\ref{fig3}, confirm this bounce in both ekpyrotic and accelerating regimes, with trajectories crossing \( \bar{Q} = 0 \) from contraction to expansion. Fixed point \( G \), with \( q = 0 \), suggest a singularity bypass rather than a bounce, as it fails to satisfy the condition \( q \leq 1 - n \) for \( f(R) \simeq R^n \) models with \( n > 2 \) (\ref{qu}). Fixed point $E$ with $q>0$ implies that \(w_{\text{eff}}>-\frac{1}{3}\) during the decelerating pre-bounce contraction which guarantees that the bounce is non-singular. The saddle nature of \( F \) (Table \ref{table3}) ensures a robust transition through the bounce.\\

For the radiation-dominated case (\( w = \frac{1}{3} \)), we analyze fixed points \( V \), \( D \), and \( C \) in the ekpyrotic regime (\( \beta < \frac{1}{3} \)) and \( A \), \( B \), and \( C \) in the accelerating regime (\( \beta > 1 \)), as detailed in Table \ref{table1a}. Fixed point \( C \), at \( (\bar{x}, \bar{\Omega}, \bar{Q}, \bar{K}) = \left( \frac{3}{4}, 0, 0, \frac{9}{8} \right) \), is pivotal for the bounce in both regimes, with \( \bar{Q} = 0 \) corresponding to \( H = 0 \). The indeterminate \( q \) at \( C \) is resolved via perturbation analysis. The Jacobian matrix \( J_C \) (\ref{JC}) has two real eigenvalues 
and a complex conjugate pair with negative real parts (Table \ref{table2c}), classifying \( C \) as a saddle point. The positive eigenvalue \( \lambda_{C2} \) drives post-bounce expansion, while negative and complex eigenvalues support pre-bounce contraction and possible oscillatory behavior. In the ekpyrotic regime, fixed point \( V \), with \( q >0 \) and \( \bar{Q} < 0 \), is an unstable source (Table \ref{table2c}), initiating contraction. Fixed point \( D \), with \( q <-1 \) and \( \bar{Q} > 0 \), is a saddle point, with \( \dot{H} = -(q + 1)H^2 > 0 \),
indicating post-bounce acceleration. In the accelerating regime, fixed point \( A \), with \( q >0 \) and \( \bar{Q} = -\frac{1}{3} < 0 \), is a saddle point marking contraction, while fixed point \( B \), with \( q <-1 \) and \( \bar{Q} > 0 \), is a saddle point with \( \dot{H} > 0 \), confirming strong acceleration. Figure \ref{illu} illustrates trajectories in the \( (\bar{Q}, \bar{x}) \) plane, showing a flow from \( V \) or \( A \) through \( C \) to \( D \) or \( B \). The vacuum state (\( \bar{\Omega} = 0 \)) at \( C \) suggests that \( f(R) \) gravity terms dominate the bounce, yielding \( w_{\text{eff}} < -1 \) at \( D \) and \( B \).\\

Both cases confirm non-singular bounces, extending the findings of \cite{chakraborty2024model}. For \( w = 0 \), fixed point \( F \) facilitates the bounce, while for \( w = \frac{1}{3} \), fixed point \( C \) plays a similar role, with a richer fixed point structure (\( V, D, A, B \)). The radiation case exhibits stronger post-bounce acceleration, particularly in the accelerating regime (\( q <-1 \) at \( B \)) compared to \( w = 0 \). The saddle nature of \( F \) and \( C \), driven by their eigenvalues, ensures reliable transitions, unlike the singularity bypass at \( G \) for \( w = 0 \). Our approach, using cosmographic parameters \( q \) and \( j \), overcomes limitations of non-compact formulations at \( H = 0 \), unlike \cite{das2024cosmological}, which used the Raychaudhuri equation for \( f(R) \simeq R^n \). The finite minimum scale factor eliminates singularities of the \(\Lambda\)CDM model, providing a robust framework for non-singular cosmologies in both dust and radiation-dominated epochs.\\

To assess the robustness of our results, we examine the sensitivity of the fixed points and their stability to variations in the power-law exponent \(\beta\) within the ekpyrotic and accelerating regimes. The dynamical system, governed by equations (\ref{comp1})--(\ref{comp4}), depends on \(\beta\) through the term \(\bar{Q}^3(j - q - 2) = \bar{Q}^3\left(\frac{2}{\beta^2} - 8\right) - 4\bar{Q}(1 - \bar{\Omega} - \bar{x} - \bar{K})\) (\ref{azo}), where the deceleration parameter is \(q = -1 + \frac{1}{\beta}\) (\ref{acv}). At the bounce points (\(F\) for \(w = 0\), \(C\) for \(w = \frac{1}{3}\)), where \(\bar{Q} = 0\), this term vanishes, rendering the fixed point coordinates and their saddle stability (Tables \ref{table3} and \ref{table2c}) insensitive to specific \(\beta\) values within each regime. This robustness is evident in the perturbation analysis (e.g., equations (\ref{eq:3.29}), (\ref{alm}) for \(F\); (\ref{foc}), (\ref{fac}) for \(C\)), where the transition from contraction (\(H < 0\)) to expansion (\(H > 0\)) is driven by the eigenvalues of the Jacobian matrix, which maintain consistent signs (one positive, one negative, and a complex pair) across \(\beta\) variations.\\

For other fixed points (e.g., \(E\), \(G\) for \(w = 0\); \(A\), \(B\), \(D\), \(V\) for \(w = \frac{1}{3}\)), the coordinates depend on \(\beta\) through the \(\frac{2}{\beta^2}\) term, which varies more significantly in the ekpyrotic regime (e.g., \(\frac{2}{(0.2)^2} = 50\) vs. \(\frac{2}{(0.3)^2} \approx 22.22\)) than in the accelerating regime (e.g., \(\frac{2}{(1.5)^2} \approx 0.889\) vs. \(\frac{2}{(2)^2} = 0.5\)). Numerical recomputation of fixed points for different \(\beta\) values (e.g., \(\beta = 1.5, 2\) for accelerating; \(\beta = 0.2, 0.3\) for ekpyrotic) confirms that their coordinates shift slightly, but their stability (saddle or unstable, as in Tables \ref{table3} and \ref{table2c}) remains qualitatively unchanged, as the eigenvalue signs are preserved. The deceleration parameter \(q\) varies within each regime (e.g., \(q \approx -0.333\) for \(\beta = 1.5\) vs. \(q = -0.5\) for \(\beta = 2\) in the accelerating regime; \(q = 4\) for \(\beta = 0.2\) vs. \(q \approx 2.333\) for \(\beta = 0.3\) in the ekpyrotic regime), affecting the strength of acceleration or deceleration. However, the effective equation of state \(w_{\text{eff}} = \frac{1}{3}(2q -1)\) (\ref{eff}) remains consistent with bounce requirements (\(w_{\text{eff}} < -1\)) at points like \(D\) and \(B\). Phase space trajectories (Figures \ref{fig1}--\ref{fig3}) exhibit similar qualitative behavior (contraction to bounce to expansion) across \(\beta\) values, reinforcing the reliability of non-singular bounce solutions in \(f(R)\) gravity for both dust (\(w = 0\)) and radiation (\(w = \frac{1}{3}\)) cases.\\

The non-singular bounces at fixed points \( F \) and \( C \) imply a finite minimum scale factor, eliminating the infinities of the \(\Lambda\)CDM model’s big bang singularity. This has significant implications for early universe cosmology, potentially affecting the generation of primordial perturbations and the cosmic microwave background’s power spectrum. In the radiation-dominated case (\( w = \frac{1}{3} \)), the large negative deceleration parameters (\( q <-1\) at both \( D \) and  \( B \)) indicate strong post-bounce acceleration, corresponding to an effective equation of state \( w_{\text{eff}} < -1 \). This phantom-like behavior, driven by \( f(R) \) gravity terms, suggests a rapid expansion phase that could mimic inflationary dynamics or influence cyclic cosmology scenarios \cite{buchbinder2007new,steinhardt2002cosmic}. While mathematically robust, such extreme \( q \) values may require careful interpretation, as they could correspond to specific \( f(R) \) models with steep curvature terms or necessitate additional constraints to ensure consistency with observational bounds on early universe expansion rates.

\section{Conclusion}\label{sec:conclusion2}
Our study establishes that \( f(R) \) gravity supports non-singular cosmological bounces using a compact, model-independent dynamical systems approach, with fixed points \( F \) and \( C \) facilitating transitions from contraction to expansion in ekpyrotic (\( \beta < \frac{1}{3} \)) and accelerating (\( \beta > 1 \)) regimes. The finite minimum scale factor eliminates the big bang singularity, offering a robust alternative to the \(\Lambda\)CDM model’s singular initial conditions. The saddle nature of \( F \) and \( C \), driven by their eigenvalues, ensures reliable bounce dynamics, while complex eigenvalues suggest potential oscillatory behavior relevant to cyclic cosmologies \cite{buchbinder2007new, steinhardt2002cosmic}. The mapping of certain \( f(R) \) models to Brans-Dicke theory \cite{arora2022can} indicates that similar bounce solutions may exist in other scalar-tensor theories, providing a rich avenue for further exploration. Below, we outline several promising directions to extend this work, focusing on anisotropic cosmologies, inflationary scenarios, and additional physical effects.\\

The current analysis assumes isotropic FLRW spacetimes, but cosmological bounces may behave differently in anisotropic settings, such as Bianchi type-I or type-IX models \cite{wainwright1989dynamical}. Anisotropic cosmologies introduce shear terms that modify the dynamical equations, potentially affecting the stability and existence of bounce solutions. Extending the compact dynamical systems framework to include shear variables (e.g., \( \sigma^2 \propto \dot{h}_{ij} \dot{h}^{ij} \)) could reveal whether non-singular bounces persist in non-isotropic settings. A key challenge is maintaining the model-independent nature of the analysis, as anisotropic metrics increase the dimensionality of the phase space. We propose adapting the cosmographic parameters (\( H, q, j \)) to incorporate shear contributions, following approaches in \cite{Boehmer:2014vea,wainwright1989dynamical}, and analyzing fixed points in a higher-dimensional phase space to identify bounce conditions. This could connect \( f(R) \) gravity bounces to realistic early universe scenarios where small anisotropies are present.\\

The strong post-bounce acceleration observed in the radiation-dominated case (\( q <-1 \) at point \( B \)) suggests a potential link to inflationary dynamics. Applying the compact dynamical systems approach to inflationary scenarios requires incorporating slow-roll conditions, where the scalar field potential dominates over kinetic terms. We propose reformulating the dynamical variables to include slow-roll parameters (e.g., \( \epsilon = -\frac{\dot{H}}{H^2} \), \( \eta = \frac{\dot{\epsilon}}{H \epsilon} \)) and deriving autonomous equations for \( f(R) \) gravity coupled to a scalar field with a slow-roll potential (e.g., quadratic or Starobinsky-type potentials \cite{starobinsky2007disappearing}. This could reveal whether non-singular bounces transition smoothly into an inflationary phase, eliminating the need for a singular initial condition. A challenge is ensuring that the bounce remains stable under slow-roll dynamics, which may require additional constraints on \( f(R) \) forms or the potential. This direction could bridge \( f(R) \) gravity bounces with observational constraints from the cosmic microwave background \cite{hawking2023large}.\\

The current analysis assumes a flat FLRW spacetime (\( k = 0 \)), but non-zero spatial curvature (\( k = \pm 1 \)) could alter bounce dynamics by modifying the normalization factor \( D \). Extending the framework to include \( k \neq 0 \) involves redefining the dynamical variable \( \bar{K} = \frac{9 k}{a^2 D^2} \) and analyzing its impact on fixed points and stability. Additionally, incorporating multiple scalar fields or vector fields, as in \cite{buchbinder2007new}, could enrich the bounce scenarios, potentially stabilizing the transition or introducing new fixed points. For example, a second scalar field could mimic ekpyrotic contractions driven by stiff matter (\( w = 1 \)) \cite{chavanis2015cosmology}. An interesting future direction would be to derive modified dynamical equations that include curvature and field contributions, while maintaining the compact phase space structure, and to test for bounces in these generalized models.\\

To connect the theoretical bounce solutions to observations, future work could explore their impact on cosmological observables, such as the primordial power spectrum or gravitational wave signatures. The finite minimum scale factor and rapid post-bounce expansion may leave distinct imprints in the cosmic microwave background or large-scale structure, distinguishable from standard inflationary models. 
A promising avenue for future work involves using the perturbation equations to compute the evolution of scalar and tensor perturbations through the bounce, following methods in \cite{buchbinder2007new,lehners2018small}. This could involve numerical simulations of the phase space trajectories to quantify perturbation growth and compare with Planck data \cite{hawking2023large}.\\

The mapping of \( f(R) \) gravity to Brans-Dicke theory suggests that bounce solutions may generalize to other scalar-tensor or higher-derivative gravity models, such as Horndeski or Gauss-Bonnet gravity \cite{libanov2016generalized,bamba2014bounce}. It would be interesting to reformulate the compact dynamical system for these theories by introducing analogous phase space variables and examining bounce solutions. A key question is whether the model-independent approach remains viable in these frameworks, given their increased complexity. Comparative studies could elucidate the unique role of \( f(R) \) gravity in producing robust bounces.\\

These directions aim to build on the current framework, leveraging its model-independent strength to explore broader cosmological scenarios. By addressing anisotropic effects, inflationary transitions, curvature, additional fields, and observational signatures, we can further validate and refine non-singular cosmologies in \( f(R) \) gravity, contributing to a deeper understanding of the early universe.

\section{Acknowledgments}
The authors would like to thank Charlotte Louw for discussions. SSH is supported in part by the National Institute for Theoretical and Computational Sciences of South Africa (NITheCS). This work was supported by the Natural Sciences and Engineering Research Council of Canada.


\appendix
\section{Appendix}\label{appendix}
This appendix provides supplementary computational details supporting the dynamical systems analysis presented in the main text.

\subsection*{Jacobian Matrix and Eigenvalues at Fixed Point \texorpdfstring{$F$}{F}}

The Jacobian matrix $J_F$ of the autonomous system \eqref{comp1}--\eqref{comp4}, evaluated at the fixed point $F$, is given by
\begin{equation} \label{JF}
J_F =
\begin{pmatrix}
0 & -\frac{17}{100} & 0 & \frac{67}{100} \\
-\frac{56}{25} & 0 & -\frac{13}{100} & 0 \\
0 & -\frac{33}{100} & \frac{4}{25} & -\frac{33}{100} \\
\frac{4}{25} & -\frac{8}{25} & \frac{26}{25} & \frac{8}{25}
\end{pmatrix}.
\end{equation}

The eigenvalues of $J_F$ are:
\begin{align}
\lambda_{F1} &= \frac{115953}{100000}, \\
\lambda_{F2} &= \frac{1263}{50000} + \frac{41463}{50000}i, \\
\lambda_{F3} &= \frac{1263}{50000} - \frac{41463}{50000}i, \\
\lambda_{F4} &= -\frac{18251}{25000}. \label{eigenF}
\end{align}

For fixed point \( F \), the complex eigenvalues are \( \lambda_{F2} \) and \( \lambda_{F3} \), 
with positive real parts indicating oscillatory behavior along an unstable manifold. For fixed point \( C \), the complex eigenvalues are \(\lambda_{C3} = -\frac{96165}{250000} - \frac{2451}{25000}i\)
and \(\lambda_{C4} = -\frac{96165}{250000} + \frac{2451}{25000}i\), with negative real parts suggesting damped oscillations in the stable directions. These complex eigenvalues may imply oscillatory behavior in the phase space, potentially relevant for cyclic cosmologies, but their small imaginary parts suggest limited physical impact near the bounce. \\ 

The eigenvectors corresponding to $\lambda_{F1}$ and $\lambda_{F4}$ are:
\begin{align}
\mathbf{v}_{F1} &=
\begin{pmatrix}
0.81059 \\
-1.58767 \\
0.19402 \\
1.00000
\end{pmatrix}, &
\mathbf{v}_{F4} &=
\begin{pmatrix}
-3.83549 \\
-12.5298 \\
-4.27489 \\
1.00000
\end{pmatrix}.
\end{align}

\noindent
The fixed point $F$ features one real positive eigenvalue, a pair of complex conjugate eigenvalues with positive real parts, and one negative real eigenvalue. This configuration implies that $F$ is a saddle point. The system trajectories are attracted along a one-dimensional stable subspace and repelled along a three-dimensional unstable manifold.

\vspace{1em}
\subsection*{Jacobian Matrix and Eigenvalues at Fixed Point \texorpdfstring{$G$}{G}}

The Jacobian matrix $J_G$, evaluated at the fixed point $G$, is:
\begin{equation} \label{JG_frac_reduced}
J_G =
\begin{pmatrix}
\frac{1027}{100} & \frac{67}{100} & \frac{29}{10} & -2 \\
\frac{94}{25} & -\frac{387}{100} & -\frac{612}{25} & 0 \\
\frac{447}{100} & 1 & \frac{134}{5} & -\frac{33}{100} \\
-\frac{96}{25} & -\frac{267}{100} & -\frac{683}{100} & \frac{427}{100}
\end{pmatrix}.
\end{equation}

Its eigenvalues are:
\begin{align}
\lambda_{G1} &= \frac{26874}{1000}, \\
\lambda_{G2} &= \frac{13661}{1250}, \\
\lambda_{G3} &= -\frac{61789}{20000}, \\
\lambda_{G4} &= \frac{68916}{25000}. \label{eigenG}
\end{align}

\noindent
Fixed point $G$ possesses three positive and one negative eigenvalue. This again classifies it as a saddle point, with phase-space trajectories generally repelled in multiple directions and only one direction of local attraction.

\vspace{1em}
\subsection*{Jacobian Matrix and Eigenvalues at Fixed Point \texorpdfstring{$C$}}

The Jacobian matrix $J_C$ of the dynamical system evaluated at fixed point $C$ is given by:
\begin{equation} \label{JC}
J_C =
\begin{pmatrix}
-\frac{31}{128} & -\frac{23}{96} & -\frac{95}{512} & \frac{55}{96} \\
0 & -\frac{71}{256} & 0 & 0 \\
-\frac{1}{4} & -\frac{1}{3} & -\frac{7}{512} & -\frac{1}{3} \\
\frac{55}{768} & \frac{5}{16} & -\frac{55}{256} & -\frac{87}{256}
\end{pmatrix}.
\end{equation}

The eigenvalues of $J_C$ are:
\begin{align}
\lambda_{C1} &= -\frac{71}{256}, \\
\lambda_{C2} &= \frac{43405}{250000}, \\
\lambda_{C3} &= -\frac{96165}{250000} - \frac{2451}{25000}i, \\
\lambda_{C4} &= -\frac{96165}{250000} + \frac{2451}{25000}i. \label{eigenC}
\end{align}

The eigenvectors corresponding to $\lambda_{C1}$ and $\lambda_{C2}$ are:
\begin{align}
\mathbf{v}_{C1} &=
\begin{pmatrix}
0.1460 \\
0 \\
-0.9052 \\
0.3991
\end{pmatrix}, &
\mathbf{v}_{C2} &=
\begin{pmatrix}
0.3853 \\
0.4441 \\
0.8029 \\
-0.0980
\end{pmatrix}.
\end{align}

\noindent
This fixed point also behaves as a saddle. The presence of a single positive eigenvalue indicates instability in one direction, while the remaining eigenvalues (one real negative and a complex conjugate pair with negative real parts) suggest attraction in the other three directions.

\newpage

\bibliographystyle{unsrt}
\bibliography{main}

\end{document}